\documentclass[twocolumn,epjc3]{svjour3}  

\RequirePackage[T1]{fontenc}

\smartqed  
\RequirePackage{graphicx}
\RequirePackage{mathptmx}      
\RequirePackage{flushend}
\RequirePackage[numbers,sort&compress]{natbib}
\RequirePackage[colorlinks,citecolor=blue,urlcolor=blue,linkcolor=blue]{hyperref}

\journalname{Eur. Phys. J. C}
\usepackage{amsmath}
\usepackage{dcolumn}
\usepackage{bm}
\usepackage{amssymb} 
\usepackage{dsfont}
\usepackage{multirow}
\usepackage[normalem]{ulem}
\usepackage[english]{babel}
\usepackage[utf8]{inputenc}
\usepackage[T1]{fontenc}
\usepackage{accents}
\usepackage{subfigure}
\usepackage{amssymb}
\usepackage{xcolor}
\usepackage[table,xcdraw]{}
\usepackage[normalem]{ulem}
\usepackage{xspace}
\usepackage{units}
\usepackage{multirow}
\usepackage{hyperref}

\newcommand{\jewel}{\textsc{Jewel}}

\newcommand{\nconst}{\ensuremath{n_\mathrm{const}}}

\RequirePackage[dvipsnames]{xcolor}

\begin{document}

\title{Quantifying vacuum-like jets in heavy-ion collisions: a Machine Learning study}



\author{Miguel Crispim Rom\~{a}o\thanksref{e1,addr1,addr2}
        \and
        Jo\~{a}o A. Gon\c{c}alves\thanksref{e2,addr2,addr3}
        \and
        Jos\'e Guilherme Milhano\thanksref{e3,addr2,addr3}
}

\thankstext{e1}{e-mail: miguel.romao@durham.ac.uk}
\thankstext{e2}{e-mail: jgoncalves@lip.pt}
\thankstext{e3}{e-mail: gmilhano@lip.pt}


\institute{Institute for Particle Physics Phenomenology, Durham University, Durham DH1 3LE, United Kingdom \label{addr1}
           \and
           LIP, Av.\ Prof.\ Gama Pinto, 2, P-1649-003 Lisboa, Portugal \label{addr2}
           \and
           Instituto Superior T\'ecnico (IST), Universidade de Lisboa, Av.\ Rovisco Pais 1, 1049-001, Lisbon, Portugal\label{addr3}
}

\date{Received: date / Accepted: date}

\maketitle

\begin{abstract}
The modification of jets by interaction with the Quark Gluon Plasma has been extensively established through the comparison of observables computed for samples of jets produced in nucleus-nucleus collisions and proton-proton collisions. 
The presence of vacuum-like jets, jets that experienced little interaction with the Quark Gluon Plasma, in the nucleus-nucleus samples dilutes the overall observed modification hindering the detailed study of the underlying physical mechanisms. 
The ability to ascertain on a jet-by-jet basis the degree of modification of a jet would be an invaluable step in overcoming this limitation.
We consider a Transformer classifier, trained on a low-level representation of jets given by the 4-momenta of all its constituents. We show that the Transformer is able to capture discriminating information not accessible to other architectures which use high-level physical observables as input. 
The Transformer allows us to identify, in the experimentally relevant case where both medium response and underlying event contamination are accounted for, a class of jets that have been unequivocally modified. Further, we perform a robust estimate of the upper bound for the fraction of jets in nucleus-nucleus collisions that are, for all purposes, indistinguishable from those produced in proton-proton collisions.
\end{abstract}

\section{Introduction}
\label{intro}

Jets result from the reconstruction of the hadronized end products of the branching of highly virtual energetic partons produced in ultra-relativistic particle collisions.
In heavy-ion collisions, the branching occurs while traversing and interacting with the Quark Gluon Plasma (QGP) created in the same collisions. 
As a  result of this interaction, properties of jets reconstructed in heavy-ion collisions differ from those in the proton-proton case. 
These modifications, globally referred to as jet quenching~\cite{Majumder:2010qh,Mehtar-Tani:2013pia,Blaizot:2015lma,Apolinario:2022vzg,Apolinario:2024equ}, have been hailed as providing a promising pathway towards characterization of the spacetime structure of the QGP \cite{Wang:2002ri,Armesto:2004pt,He:2020iow,Barata:2022krd,Barata:2023qds,Barata:2023zqg,Barata:2024bqp,Bahder:2024jpa,Ke:2024emw,Barata:2025bgl,Apolinario:2017sob,Apolinario:2020uvt,Attems:2022ubu,Brewer:2025wol,Apolinario:2024hsm}.
Jet quenching has been unequivocally established from a variety of global and substructure jet observables, whose experimentally measured distributions for jet samples reconstructed in heavy ion collisions differ from those measured in proton-proton collisions \cite{Connors:2017ptx,Cunqueiro:2021wls,Apolinario:2022vzg,Apolinario:2024equ}.

The extent to which each individual jet is modified is known to depend on several factors, including the multiplicity of its vacuum-like fragmentation \cite{Milhano:2015mng,Rajagopal:2016uip,Casalderrey-Solana:2018wrw,Escobedo:2016jbm}, the amount of matter traversed, and the gradients of that matter \cite{Barata:2022krd,Barata:2023qds,Barata:2023zqg,Barata:2024bqp,Bahder:2024jpa,Ke:2024emw,Barata:2025bgl}.
A standard measure of the modification of a jet is the comparison of its reconstructed total transverse momentum with a proxy for the transverse momentum of the parton from which the jet originated, be it the transverse momentum of an electroweak boson in events where it is produced back-to-back with a jet, or in the more abundant dijet case through an estimate of average energy loss \cite{Brewer:2018dfs,Takacs:2021bpv,Apolinario:2024apr,Andres:2024hdd}. This measure disregards the possibility of jets being modified by interaction with QGP without significant energy loss \cite{Apolinario:2025fwd}. 

To devise a novel measure that is sensitive to the modification of properties of a jet beyond its total energy amounts to establishing whether a modified jet contains enough information, without reference to any external proxy, to be  distinguishable from a jet produced in the absence of QGP. In the context of Machine Learning, one can ask whether a machine can learn to make such a distinction. Several studies \cite{Du:2020pmp,Du:2021pqa,Apolinario:2021olp,Lai:2021ckt,Liu:2022hzd, CrispimRomao:2023ssj,Qureshi:2024ceh,ArrudaGoncalves:2025wtb,Li:2025tqr,Goncalves:2025asw,Wu:2025kqh}, including by some of us \cite{Apolinario:2021olp,CrispimRomao:2023ssj,ArrudaGoncalves:2025wtb,Goncalves:2025asw}, have addressed this point. In particular, 
it was established \cite{CrispimRomao:2023ssj} that such discrimination is possible on the basis of combinations of standard observables, with pairs of observables saturating the discrimination power of the large set of observables considered in that work.

Here, we address this task by considering a low-level representation of jets, i.e. the list of the 4-momenta of all its constituents, to train a Transformer classifier~\cite{vaswani2017attention}.
This allows us to ask and answer whether there is discriminating information within the jet that is not captured by the standard high-level observables considered in \cite{CrispimRomao:2023ssj}. 
Transformers arise as the natural architecture to seek such answers. Their versatility and unparalleled performance for sequence-type data, which low-level jet data can be represented as, has taken the High Energy Physics (HEP) community by storm~\cite{Mikuni:2021pou,Qu:2022mxj,Builtjes:2022usj,Kach:2022uzq,DiBello:2022iwf,Finke:2023veq,Butter:2023fov,He:2023cfc,Hammad:2023sbd,Spinner:2024hjm,Brehmer:2024yqw,Spinner:2025prg,Gambhir:2025xim}.
They consistently match or exceed the performance of established architectures across a range of tasks using only low-level data, from jet tagging to generative modelling.

In turn, this allows us to establish an upper bound on the fraction of jets produced in PbPb collisions that are, for all purposes, indistinguishable from their vacuum counterparts produced in proton proton collisions.

The paper is organized as follows: Section~\ref{sec:datasim} describes the generated datasets;  Section~\ref{sec:transformer} introduces the transformer classifier;  Section~\ref{sec:results} shows and discusses our classification results;  Section~\ref{sec:vacinmed} provides an estimate of the fraction of vacuum-like jets in jets produced in PbPb collisions; and we conclude in  Section~\ref{sec:conclusions}. \ref{app:observables} and \ref{app:hp-opt} detail, respectively, the high-level observables 
 used in \cite{CrispimRomao:2023ssj} and the optimisation of the transformer hyperparameters.

\section{Data Simulation}
\label{sec:datasim}

This study is based on di-jet samples produced by the Monte-Carlo event generator \jewel~2.3 \cite{Zapp:2013vla}. 
Two datasets, each including both the proton-proton (vacuum) and Pb-Pb (medi\-um) cases, were prepared. 
The dataset referred to as \textit{signal only} (SO) does not include any effects due to QGP response, nor effects due to contamination from imperfectly subtracted underlying-event.
The \textit{medium response and underlying event} (MR-UE) dataset was prepared following the procedure introduced in \cite{ArrudaGoncalves:2025wtb}, where full details can be found. In this case, the PbPb sample includes effects due to QGP response as modelled by \jewel\ -- including the \jewel\ specific subtraction \cite{Milhano:2022kzx} which avoids double counting of underlying event contributions -- is embedded in a realistic heavy-ion underlying event which is subsequently subtracted using standard methods. The proton-proton sample is also embedded in a realistic heavy-ion underlying event and subtraction is carried out analogously to the PbPb case. This proton-proton baseline corresponds to the physically relevant case where jets are not modified by interaction with QGP, but display modifications entirely due to underlying-event contamination.

All samples were generated at hadron level for $\sqrt{s} = 5.02$~TeV with a hard matrix element lower transverse momentum cut-off of $50$~GeV and the generation spectrum was re-weighted by $p_T^5$ as to oversample the large $p_T$ region. The resulting event weights were used throughout the study. For the PbPb samples, the QGP was generated according to the simple parametrization described in \cite{Zapp:2013zya} with $\tau_i = 0.4$~fm/$c$, $T_i = 590$~MeV, $T_c = 170$~MeV, centrality $0-10\%$, and covering $|\eta| < 4$.

Jets were reconstructed from all particles within $\eta_{\textsf{part}}< 2.5$ using the anti-$k_T$ clustering algorithm \cite{Cacciari:2008gp}, with $R=0.4$, as implemented in \texttt{fastjet}~\cite{Cacciari:2011ma}. 
Only jets with $p_T \in [80, 230]$~GeV, $\eta_{\textsf{jet}} <2.0$, and with at least two constituents were retained in the analysis. 
The approximate number of reconstructed jets is $\mathcal{O}(1.7\times 10^6)$ for SO ($\mathcal{O}(10^6)$ vacuum, $\mathcal{O}(7\times10^5)$ medium) and $\mathcal{O}(1.9\times 10^6)$ for MR-UE ($\mathcal{O}(10^6)$ vacuum, $\mathcal{O}(9\times10^5)$ medium).

Once reconstructed, a jet is a collection of the 4-mo\-men\-ta of its constituents.
For this work, we chose to represent the 4-momentum of each constituent by its rapidity and azimuth $(\Delta \eta, \Delta \phi)$ relative to the jet axis, its transverse momentum $p_T$, and its mass $m$.\footnote{For the mass we used the on-shell mass associated with the particle ID, not the one that \texttt{fastjet} derives, as this can lead to non-physical masses after the subtraction steps. More importantly, we observed that this leads to an artificial and non-physical large discrimination power for the MR-UE sample due to the \jewel-specific subtraction prescription.}  
A single jet is then the set $\{( \log(p_{T,i}/(1 GeV)), m_i,\Delta \eta_i, \Delta \phi_i)\}$ with $i=1,\dots,M$ where $M$ is the number of jet constituents. The use of the logarithm of the transverse momentum ensures that the values are nominally similar across all constituents, which is an important requisite for the neural network. While this representation cannot be obtained by a Lorentz transformation  of a physical 4-momenta, it retains all the same information. \footnote{We leave for future work how different handling of the 4-momenta, for example by using Lorentz algebra-aware architectures~\cite{Spinner:2024hjm,Brehmer:2024yqw,Spinner:2025prg}, can impact the analysis presented in this work.} 

Each jet is represented as a matrix $\mathbf{X} \in \mathbb{R}^{M\times 4}$ with $\mathbf{X} = X_{i\alpha}$, where $i=1,\dots,M$ runs over the jet $M$ constituents, and $\alpha=1,\dots,4$ runs over the four physical quantities. 

For comparison purposes, we also computed the high-level observables, such as jet substructure observables, discussed in our previous work~\cite{CrispimRomao:2023ssj}. This comparison will allow us to assess how much information inside the jet is not being captured by observables usually employed in phenomenological and experimental studies.
The dataset used for the machine learning training and the analysis presented herein is made available here~\cite{zenodo}.


\section{Transformer and methodology}
\label{sec:transformer}

In previous work~\cite{Apolinario:2021olp,CrispimRomao:2023ssj,ArrudaGoncalves:2025wtb}, we considered different machine learning classification algorithms which embodied different \emph{inductive biases}, i.e. assumptions on how the data representation encodes the relevant information.
For example, for jets represented as images, Convolutional Neural Networks were used, thus assuming a compositional hierarchy from local receptive fields to a global structure arising from their successive overlaps. 
In this work, we employ the \emph{Transformer} block, initially proposed for Natural Language Processing~\cite{vaswani2017attention} as a sequence-to-sequence model, which provides a \emph{minimal inductive bias} as it intakes \emph{sets} of vectors, with no other data assumption. We now review this architecture with a notation more customary to physicists.

The Transformer classifier used in this work, $\mathcal{T}$, can be seen as the composition of four different modules: initial input embedding, $I$; $N_T$ Transformer blocks, $T$; pooling stage, $P$; and a classifier, $C$:
\begin{equation}\label{eq:transformer}
    \mathcal{T}(\mathbf{X}) = C \circ \ P \circ T^{N_T} \circ I(\mathbf{X})  \ ,
\end{equation}
where $\mathbf{X}$ is a jet represented as the set of the 4-momenta of its constituents, and each module has its own trainable weights and biases, and $T^{N_T} = T \circ T \circ \dots$ represents a $N_T$ stack of $T$ blocks. The input embedding, $I$, is an affine projection that maps the momenta of the jet constituents to a higher dimensional space, i.e.  I : $\mathbb{R}^{4} \to \mathbb{R}^{d_R}$ acting on the second index of $\mathbf{X}_{i\alpha}$. We will represent this transformation as
\begin{equation}
    I(\mathbf{X}) = \mathbf{X} w_{I} + b_{I}\, ,
    \label{eq:emb}
\end{equation}
with weights $w_I\in \mathbb{R}^{4\times d_R}$ and bias $b_I \in \mathbb{R}^{M\times d_R}$, where $d_R$ is the \emph{representation} dimension that we will keep the same for every module of Eq.~\ref{eq:transformer} for simplicity. We note that $b_I$ is the same for all jet constituents, i.e. its lines are all the same $b_{I,i\alpha}=\delta_{ii} b_{I,\beta}$ with $b_{I,\alpha} \in \mathbb{R}^{d_R}$ and its representation as a $M\times d_R$ matrix is for convenience.

This step allows the neural network to operate in a higher dimensional space, where it can abstract more complicated representations of the jet. The next module, the Transformer block, is the main ingredient of this architecture and is itself composed of two steps: the multihead self-attention, $MHA$, and a residual feed-forward, $FF$, block. The $MHA$ module reads
\begin{align}
    MHA(I) & = (\text{Concat}(\text{Att}(I)_{1},\dots, \text{Att}(I)_{N_h})) W_O\, ,  \label{eq:MHA} \\ 
    \text{Att}(I)_a & = \text{Softmax}\left(\frac{1}{\sqrt{d_R}} W_{Q,a}(I) W_{K,a}(I)^T \right) W_{V,a}(I) \label{eq:Att} \, ,
\end{align}
where $I=I(\mathbf{X})$, Concat is the concatenation operation along the second index, $\text{Softmax}(A)_{ij} = \exp(A_{ij})/\sum_k \exp(A_{ik})$ normalises the entries along the second index, Att is a single self-attention head, $N_h$ is the number of heads, $W_{V,Q,K,O}$ represent affine transformations of the form $W(A) = A w +  b$ with weights $w_{V,Q,K}\in\mathbb{R}^{d_R \times d_R}$, $w_O \in \mathbb{R}^{N_h d_R \times d_R}$, and biases $b_{V,Q,K,O}\in\mathbb{R}^{M\times d_R}$ where again the lines of the biases matrices are all the same just as in Eq.~\ref{eq:emb}.\footnote{In practice, each attention head, Att$(I)_a$, acts only on a $d_R/N_h$ slice of the representation, such that the concatenation returns a $d_R$ representation, making $W_O$ unnecessary. Nonetheless, this is how the mechanism was first presented in~\cite{vaswani2017attention}.} The subscripts $V$, $Q$, $K$, $O$ have historical names that are not relevant for this discussion, but are kept for bookkeeping. More importantly we note that the $MHA$ module learns a dense adjacency matrix between all constituents of a jet, i.e. $W_Q(I) W_K(I)^T\in \mathbb{R}^{M\times M}$, which is then transformed non-linearly by the Softmax function. The resulting matrix has normalised rows and is then contracted with $W_V(I)$ (often called Scaled Dot-Product), producing a new representation of the jet where each constituent now has the information of the whole jet.\footnote{The fact that $W_Q(I) W_K(I)^T\in \mathbb{R}^{M\times M}$ means that the new jet representation is affected by the ``context'' of the rest of the jet. Recently there has been a great deal of research around the idea that Transformer capacity to outperform other architectures for similar tasks is due to this ``in-context learning'' that Transformers are capable of performing~\cite{garg2022can,bai2023transformers}.} The rest of the Transformer block involves a residual connection and a feed-forward module such as
\begin{align}
    res(I) & = \text{LayerNorm}(MHA(I) + I)\, , \\
    T(I) & = \text{LayerNorm}(FF(res(I))+res(I)) \, ,
\end{align}
where LayerNorm normalises the inputs over the last index. i.e. the representation, and $FF$ is a multi-layer perceptron with dimension $d_R$ with one hidden layer with the ReLu non-linear activation function. The residual allows for deeper architectures as it prevents vanishing gradient problems during training and intuitively makes the transformer block $T$ act as an incremental mapping around its inputs.

Two features of the Transformer block are worth discussion. The first one, is that it preserves the dimensionality of its inputs, i.e. $T : \mathbb{R}^{M \times d_R} \to \mathbb{R}^{M \times d_R}$, so one can stack $N_T$ $T$ blocks to form a deeper neural network that learns sequentially more complex relations between the jet constituents. Second, $T$ is covariant under permutations of the jet constituents. If we consider a permutation matrix $\Pi$ that permutes the constituents of the jet: $\Pi_{\pi(i) i}\mathbf{X}_{i\alpha} = \mathbf{X}_{\pi(i)\alpha}$, where $\pi$ is some permutation over the index $i$, then $T(I(\Pi\,\mathbf{X}))=\Pi\, T(I(\mathbf{X}))$. This second feature is why Transformers operate optimally over sets of jet constituents.

The fact that the output of the Transformer block has the same dimensionality as its inputs means that the output representation has different dimensions for different jets, as $M$ (the number of constituents) varies on a jet-by-jet basis. In order to perform a classification task over this $M \times d_R$ representation, we need to \emph{pool} along the first index, i.e. over the jet constituents. We do this by defining a pooling module, $P$, which has the sole purpose of mapping the whole jet to a fixed length representation, $P: \mathbb{R}^{M\times d_R}\to\mathbb{R}^{d_R}$.
We also notice that whereas $T$ is permutation covariant, $P(T)$ is permutation invariant, and consequently the whole Transformer classifier $\mathcal{T}(\mathbf{X})$ will be permutation invariant. 
In this work we will let our hyperparameter optimisation loop choose between different pooling operations: mean, sum, max, and attention pooling.

Having pooled the whole jet representation into a vector of fixed size, we can perform the classification using $C$, a feed-forward multilayer perceptron with $M_h$ number of hidden layers of size $d_R$, which maps the representation $P(T)$ into the probability of a jet being of class one, which we define as being the medium sample, i.e. $C: \mathbb{R}^{d_R} \to [0,1]$.

We implemented the Transformer classifier in \texttt{pytorch} \cite{paszke2019pytorch}, using \texttt{lightning} \cite{falcon2019pytorch}. 
Complementarily, a gradient boosted classifier (BDT), implemented with XGBoost~\cite{Chen:2016:XST:2939672.2939785}, was trained on the same task using the high level observables considered in \cite{CrispimRomao:2023ssj}. 
This set of observables includes only observables that return a a single value per jet, spanning both global jet properties and jet substructure. 
In \ref{app:observables} we provide explicit details of the observables that are relevant for the discussion in this work. The inclusion of this classifier will allow us to compare the Transformer classifier performance on low-level jet constituent data against high-level jet data. 
The details of the hyperparameter optimisation for both algorithms are presented in~\ref{app:hp-opt}

\section{Classification Results}
\label{sec:results}

We now discuss the discrimination performance of the classifiers trained on separating vacuum (proton-proton) from medium (Pb-Pb) samples. We start by discussing the SO case, where neither QGP response nor UE contamination are accounted for. This will allow us to focus the discrimination task on modifications of the jet fragmentation history alone.

\begin{figure}[t]
    \centering
    \includegraphics[scale=0.4]{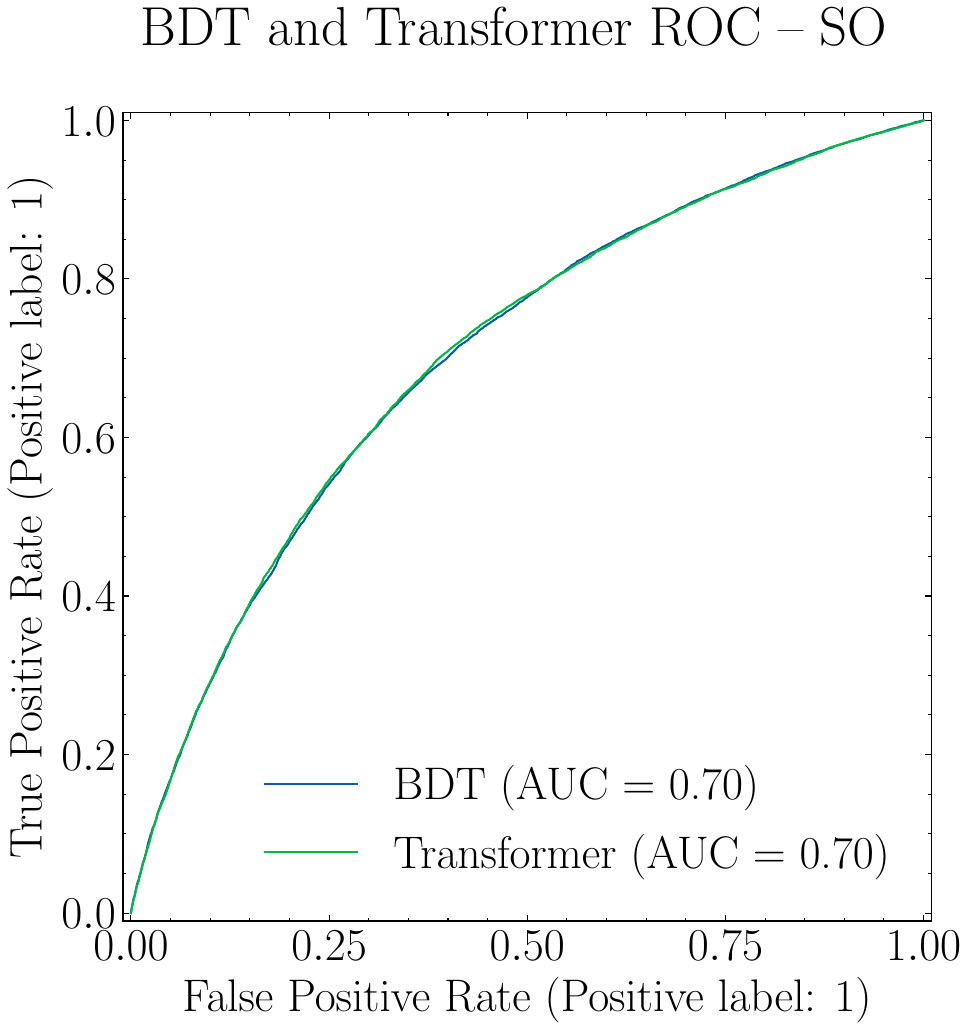}
    \caption{ROC curves and respective AUC for the BDT (Blue) and Transformer (Green) for SO scenario.}
    \label{fig:so-roc}
\end{figure}

In Fig.~\ref{fig:so-roc} we show the Receiver Operator Characteristic (ROC) curve and the Area Under the Curve (AUC), for both the Transformer and the BDT classifiers trained on the SO dataset. A very striking observation is the almost complete overlap between the two, with each ROC curve yielding an AUC of $0.70$ at two significant digits. 
This result suggests that our Transformer is not capturing any further information beyond that encoded in the high-level observables studied in~\cite{CrispimRomao:2023ssj}.

\begin{figure*}[t]
    \centering
    \includegraphics[scale=0.4]{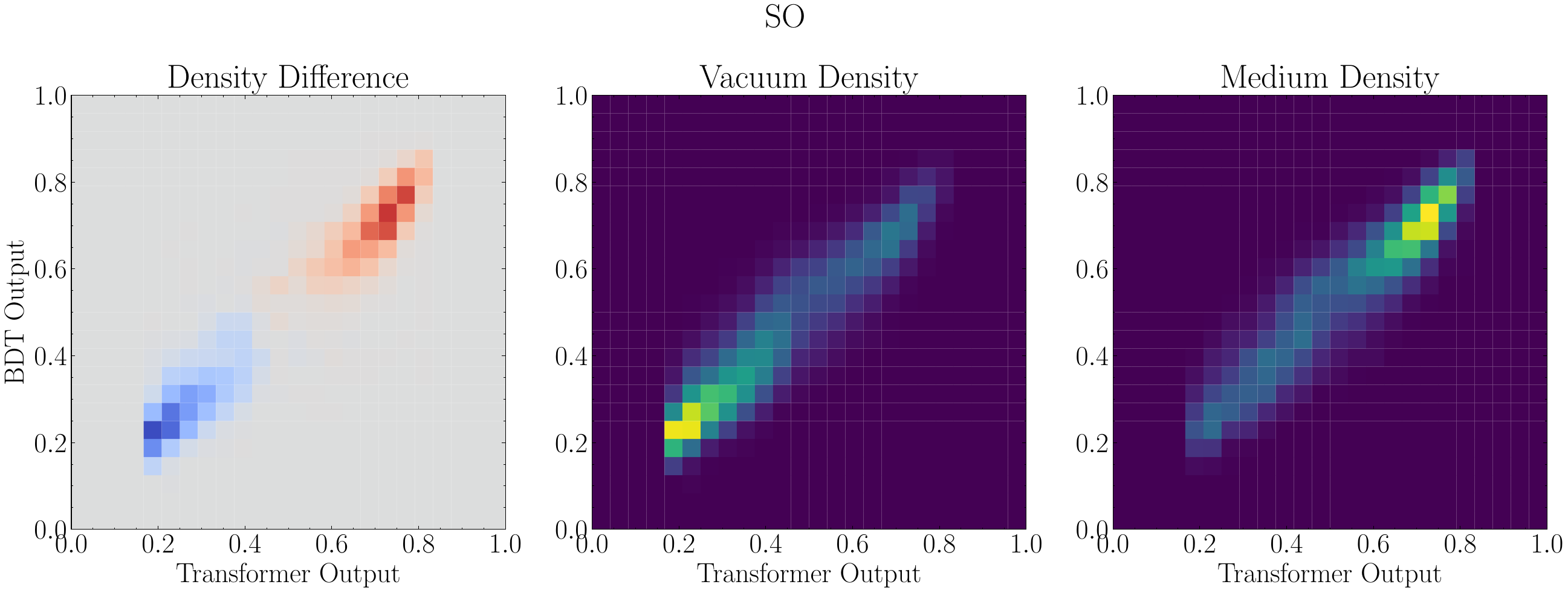}
    \caption{BDT and Transformer predictions for the SO scenario. Left: Difference of the 2d histogram density between Vacuum and Medium samples, with Blue (Red) showing a greater density of Vacuum (Medium) density. Middle: The 2d histogram density for the Vacuum sample. Right: The 2d histogram density for the Medium sample.}
    \label{fig:so-transformer-vs-bdt}
\end{figure*}

The virtually complete overlap of the ROC curve implies that both discriminants share the same quantiles, suggesting a correlation between them. This is further demonstrated in Fig~\ref{fig:so-transformer-vs-bdt} where we show the distributions of the Transformer and BDT outputs for the vacuum and medium samples of the SO dataset. The correlation is nearly perfect, with both discriminants agreeing to high degree of correlation what is a vacuum and a medium jet. This reinforces the interpretation that there is no novel information inside the jet to be captured by the Transformer that the high-level discriminants have not already used to determine  whether the jet has experienced any modification by interaction with the QGP medium.

We now turn to the MR-UE dataset, where the inclusion of both medium response and confounding effects from the imperfectly subtracted underlying event presents a scenario closer to experimental reality. While, on the one hand,  the embedding in  and subtraction of the underlying event might hide the `signal', making it harder to discriminate between vacuum and medium jets, the presence of medium response within the reconstructed jets should make medium jets less like vacuum (i.e., unmodified) jets \cite{Goncalves:2025fpf} and facilitate the discrimination task.

\begin{figure}[t]
    \centering
    \includegraphics[scale=0.4]{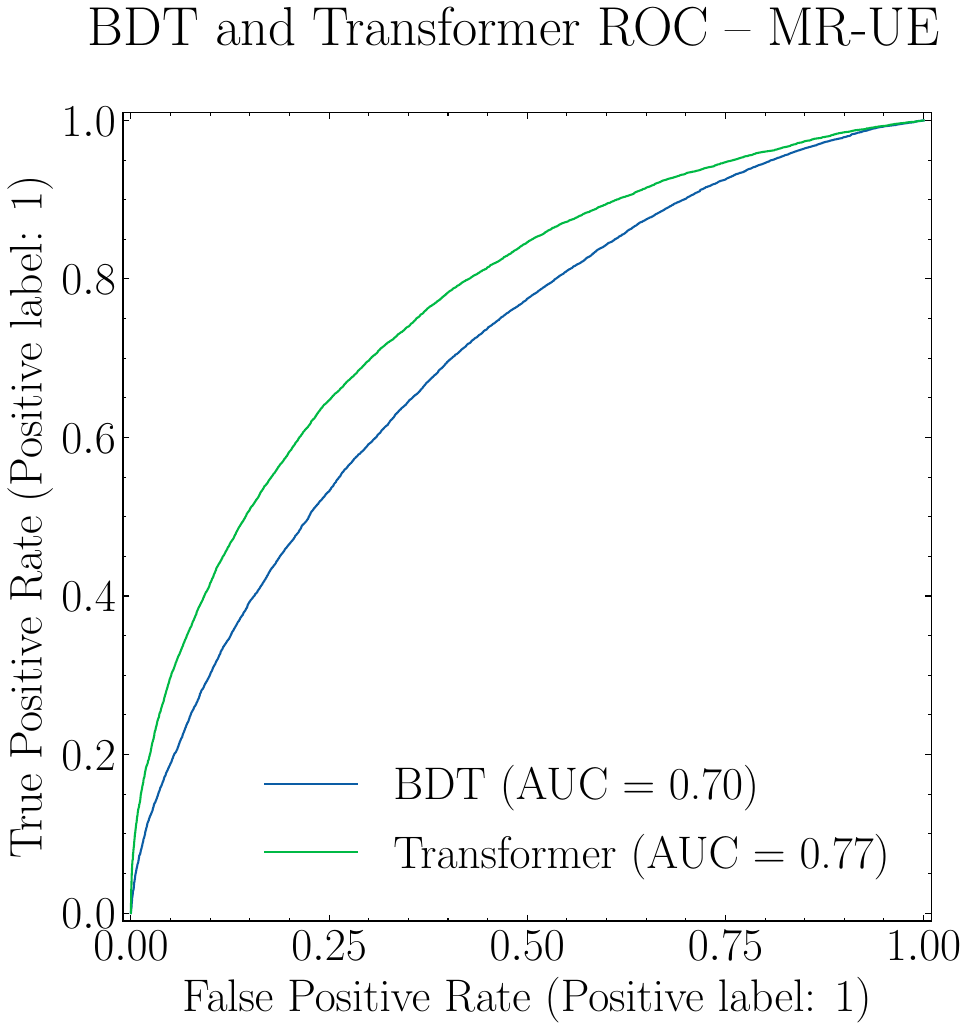}
    \caption{ROC curves and respective AUC for the BDT (Blue) and Transformer (Green) for MR-UE scenario.}
    \label{fig:mr-ue-roc}
\end{figure}

In Fig.~\ref{fig:mr-ue-roc} we show the ROC and respective AUC for the Transformer and the BDT classifiers. Interestingly, while the ROC AUC for the BDT is similar to the SO case, at $0.70$, the picture is fundamentally different for the Transformer, yielding a ROC AUC of $0.77$. This provides strong evidence that there is extra discriminating information inside the jet arising from the medium response not captured by the high-level observables that the BDT had access to during training.

\begin{figure*}[t]
    \centering
    \includegraphics[scale=0.4]{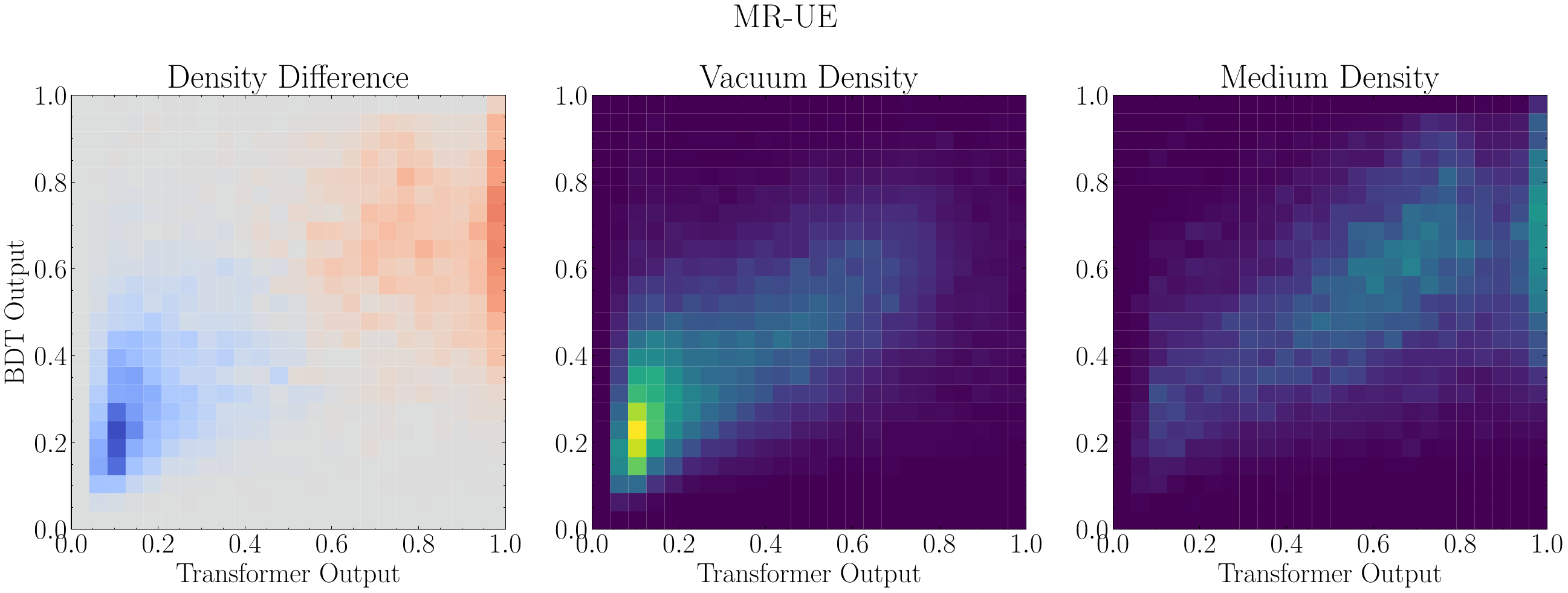}
    \caption{BDT and Transformer predictions for the MR-UE scenario. Left: Difference of the 2d histogram density between Vacuum and Medium samples, with Blue (Red) showing a greater density of Vacuum (Medium) density. Middle: The 2d histogram density for the Vacuum sample. Right: The 2d histogram density for the Medium sample.}
    \label{fig:mr-ue-transformer-vs-bdt}
\end{figure*}

Since the discriminating power of the Transformer and the BDT on the MR-UE dataset is so different, these discriminants cannot be strongly correlated. In Fig.~\ref{fig:mr-ue-transformer-vs-bdt} we show the correlation between the output distributions of these two discriminants for both samples in the MR-UE dataset. The most interesting feature of these plots is the accumulation of medium jets with Transformer prediction very close to $1.0$ (i.e., classified as medium), while the BDT prediction is evenly spread out in $[0.5,1.0]$. The second interesting observation is when we look at the remaining distributions if we ignore this modality. In this case, we observe that the Transformer and the BDT distributions for the vacuum sample and the remainder of the medium sample are highly correlated, similarly to how they were for the SO case. This suggests that in this regime, the medium response is not meaningfully modifying the jets, and we recover fragmentation patters that are similar to the SO case. Conversely, our methodology suggests that it is possible to isolate with a degree of purity jets that have medium response modifications.

Given the difference in discrimination between the Transformer and the BDT for the MR-UE scenario, it is important to assess what could the Transformer be learning that the BDT could not. While Neural Networks are often perceived as black-box functions allowing little interpretability of their predictions, the explicit expressions for the operations carried out inside the Transformer allow us to gather significant insight. 

We start with the crucial ingredient of the Transformer architecture: the dense adjacency matrix produced by the attention mechanism Eq.~\ref{eq:Att}. We see that for the first transformer block we have
\begin{align}
    W_{Q,K} (I(\mathbf{X})) & = (  \mathbf{X} w_I +  b_I) w_{Q,K} + b_{Q,K} \nonumber  \\
            & =   \mathbf{X} \tilde w_{Q,K} +  \tilde b_{Q,K} \ ,
\end{align}
where $\tilde w_{Q,K} = w_I w_{Q,K} $, $\tilde{b}_{Q,K}=b_I w_{Q,K} + b_{Q,K}$, which is nothing else than the fact that the composition of two affine transformations is an affine transformation. This leads to the adjacency matrix
\begin{align}
    W_Q(I)W_K(I)^T  =& ( \mathbf{X} \tilde w_Q + \tilde b_Q) ( \mathbf{X} \tilde w_K + \tilde b_K)^T \nonumber\\
                    =&  \mathbf{X} \tilde w_Q \tilde w_K^T \mathbf{X}^T  +  \mathbf{X} \tilde w_Q \tilde b^T_K
                    +\tilde b_Q \tilde w^T_K \mathbf{X}^T  + \tilde b_Q \tilde b_K^T \ ,
\end{align}
or, in other words, we obtain a second order polynomial in the jet constituents. We also notice that the first term is closely related to the variance matrix between jet constituents and it can be seen as the learnable pair-wise measurement of relation between the jet constituents. Next, we expand the Softmax operation, which for small arguments reads
\begin{equation*}
    Softmax(A_{ij}) \simeq (1 + A_{ij})/\sum_{k=1}^{M}(1 + A_{ik}) \simeq (1 + A_{ij})/M \, .  
\end{equation*}
This represents a re-scalling of the entries of this matrix, and therefore the output of the MHA block is schematically
\begin{equation}
    MHA(I) \simeq P^2(I) I = P^3(\mathbf{X}) \ ,  
\end{equation}
where, to first approximation, $P^i$ is an $i^{th}$ order polynomial of the $\mathbf{X}$ entries. Now we note that the outputs of $MHA$ take values close to $0$ due to the smallness of $w$ and the inputs. Therefore, the LayerNorm operation is another $O(1)$ rescalling, and we can approximate the entire first Transformer block, $T(I(\mathbf{X}))$, as
\begin{align}
    T(I) \simeq I+ MHA(I) + FF(I+MHA(I))  \ .
\end{align}
Collecting the terms in powers of $I$ dependence, we have
\begin{align}
    T(I(\mathbf{X})) \simeq P^1(I(\mathbf{X})) + P^2(I(\mathbf{X})) \, I(\mathbf{X}) \simeq P^3(\mathbf{X}) \ ,
\end{align}
and therefore, at leading order, each Transformer block outputs a cubic polynomial of its inputs. We notice, however, that the true expressive power of the Transformer block is nonetheless greater, as beyond leading order the coefficients of the polynomial are non-linear functions of the entries themselves due to the Softmax and the non-linear activation function of the $FF$ step.

In this work, the best hyperparameters for the Transformer suggested a maximum needed number of stacked Transformer blocks for both datasets at $N_T = 3$. This means that, intuitively, the transformer classifiers studied in this work have operated on a polynomial of approximate degree $3^3=27$, which is considerably higher than the order needed to produce the high level observables used to train the BDT as detailed in~\cite{CrispimRomao:2023ssj}. This could explain why the BDT was not able to capture the medium response in the MR-UE case and hint at jet substructure observables involving correlations of high order between jet constituents being needed in order to capture medium response. However, for the SO dataset we obtained a high correlation between the Transformer and the BDT, which suggests that it is likely that the optimisation loop for the Transformer `exaggerated' the need for a deeper architecture when a shallower one could have worked as well. This is expected to a certain extent, as machine learning workflows tend to produce large models that are trained with regularisation that reduce the effective functional representational capacity -- in our case dropout and early stop. Therefore, the degree $27$ should be read as an approximation to the upper bound on the required degree and not as the necessary degree of a polynomial over the jet constituents to perform this classification task.

\section{Estimating the Fraction of Vacuum in Medium}
\label{sec:vacinmed}

The analysis we carried out in the preceding section highlights the difficulty in isolating jets modified by the QGP as vacuum-like jets still represent an irreducible component of the medium sample. 
This is evident by the non-existence of a classification regime where the False Positive Rate can be brought to zero while keeping a non-zero True Positive Rate, which would be the required cut to generate a pure sample of modified jets.

A question that arises is how much of the medium sample is effectively vacuum-like, i.e. not modified enough, or at all, by interaction with the QGP. 
A crude estimate of the fraction of jets that is not modified enough by QGP can be made on the basis on jet suppression alone. Given the steepness of the jet spectrum, jets in PbPb collisions (the medium sample) with a given $p_T$ are roughly those that either did not lose any energy or lost very little \cite{Brewer:2018dfs,Apolinario:2024apr}. Jets that would have lost a lot of energy, that is that had a substantially higher initial energy, are suppressed by the steepness of the jet spectrum and contribute only marginally to the number of jets found at a given $p_T$. As such, the value of the jet nuclear modification factor $R_{AA}$ (for the $0-10\%$ centrality class $R_{AA} \sim 0.4\div 0.5$ for all considered $p_T$) is a rough upper bound for the fraction of jets that are vacuum-like.

More detailed information like that contained both in the observables we considered for BDT training and in the full set of 4-momenta we used as input for the Transformer should lead to an improved bound. It is conceivable that jets are modified in ways, e.g. in their substructure and/or correlations among constituents, without an energy loss large enough to modify the $R_{AA}$. 

Here we  will use the discriminants discussed above to derive upper limits on the fraction of vacuum-like jets in the medium sample, or, more colloquially, the fraction of vacuum in the medium. We use a simplified form of topic modelling, similar to that used to discriminate between quark and gluon initiated jets \cite{Metodiev:2018ftz,Komiske:2018vkc,ATLAS:2019rqw,Komiske:2022vxg,Dolan:2023abg,Duan:2025lvi}, towards extraction of the strong coupling \cite{LeBlanc:2022bwd}, for four-top searches \cite{Alvarez:2019knh}, and to probe the different interaction of quarks and gluons with QGP \cite{Brewer:2020och}. 

For any discriminant $\mathcal{D}$ we define its distribution for the medium sample as $\mathcal{M}_{\mathcal{D}}(x)$, where $x$ is valued over the domain of $\mathcal{D}$. The shape of this discriminant will change if the amount of modifications to the jet by the medium becomes more or less pronounced, for example by the medium temperature, density, etc. We formally parametrise, but otherwise do not specify, the quenching-inducing quantities as $\{q\}$. Therefore, for a given physical system that produces QGP medium through which the jets traverse and can interact with, we have
\begin{equation}
    \mathcal{M}_{\mathcal{D}}(x) = \mathcal{M}_{\mathcal{D}}(x|\{q\}) \ ,
\end{equation}
which is just a way of formally parametrising the `amount' of quenching that the medium can produce.

For a SO-like scenario, jet modifications produce a distributional migration of the different observables along an unidimensional trajectory \cite{CrispimRomao:2023ssj}. Therefore, we take the simplification that there is effectively only one degree of freedom parametrising quenching, i.e. $\{q\} \to q$. With this framing, we can define the unquenched jets, $u$, as the contribution to $\mathcal{M}_{\mathcal{D}}$ that has not experienced any modification. In other words, the unquenched jets are those that would have been produced in the limit $q\to 0$, which leads to
\begin{equation}
    \mathcal{M}_{\mathcal{D}}(x|q) = f_q \mathcal{M}_{\mathcal{D}}(x| q \neq 0) + f_{u} \mathcal{M}_{\mathcal{D}}(x|q =0)\, ,
\end{equation}
where the relative fractions respect $f_q + f_u = 1$. It is clear now that we should identify the unquenched distribution to the one we would obtain in the absence of medium, i.e. the vacuum, $\mathcal{M}_{\mathcal{D}}(x|q=0)\simeq\mathcal{V}_{\mathcal{D}}(x)$, and the distribution of the quenched jets as $\mathcal{Q}_{\mathcal{D}}(x)=\mathcal{M}_{\mathcal{D}}(x| q \neq 0)$. When considering experimental data, this only holds when PbPb underlying event contamination is also accounted for in the pp sample \cite{ArrudaGoncalves:2025wtb}. We have then
\begin{equation}
    \mathcal{M}_{\mathcal{D}}(x) \simeq f_q \mathcal{Q}_{\mathcal{D}}(x) + f_{v} \mathcal{V}_{\mathcal{D}}(x) \ ,
\end{equation}
where we renamed $f_u=f_v$ for consistency.

Consider now the histogram approximation of $\mathcal{D}(x)$ using $N$ bins $\mathcal{H}_{\mathcal{D},N}=\{h_{\mathcal{D}}[k]\}$, with $k=1,\dots,N$ where $h_{\mathcal{D}}[k]$ are the bin heights and with edges over the domain of $x$. Each bin height, $h_{\mathcal{D}}[k]$, results from two contributions: the vacuum, $h_{\mathcal{D},v}[k]$, and the quenching, $h_{\mathcal{D},q}[k]$, such that
\begin{equation}
    h_{\mathcal{D}}[k] = f_{q} h_{\mathcal{D},q}[k] + f_{v} h_{\mathcal{D},v}[k] \ .
\end{equation}
Given that both $\mathcal{Q}$ and $\mathcal{V}$ are valid distributions generated by different physical phenomena, they need to be non-negative. Immediately, this means that
\begin{equation}
    h_{\mathcal{D},q}[k] = \frac{h_{\mathcal{D}}[k]-f_v h_{\mathcal{D},v}[k]}{f_q} = \frac{h_{\mathcal{D}}[k]-f_v h_{\mathcal{D},v}[k]}{1 - f_v} \geq 0
\end{equation}
for all bins as long as $f_q \neq 0$, which is a reasonable assumption. Therefore, we can set a maximum bound on $f_v$, the fraction of vacuum, computed using the discriminant $\mathcal{D}$ by saturating the inequality, i.e.
\begin{equation}\label{eq:f_max}
    f^{\text{max}}_{\mathcal{D},v} = \min_{k:h_{\mathcal{D},v}[k]>0} \frac{h_{\mathcal{D}[k]}}{h_{\mathcal{D},v}[k]} \ .
\end{equation}

The dependence on the discriminant is crucial and will significantly affect our upper bound estimate. To see this, we notice that if $h_{\mathcal{D},v}$ and $h_{\mathcal{D},q}$ are highly overlapping, then they could both account for the shape of $h_{\mathcal{D}}$. This would make any value of $f_v$ admissible and the minimisation step could overestimate the quantity of $f_v$ as being the unique component. Consequently, the upper bound on the fraction of vacuum in the medium sample is more stringent for discriminants that have more pronounced differences for vacuum and quenched jets. Therefore, the machine learning discriminants developed in the previous section are expected to produce the lower upper bounds for the fraction of vacuum in the medium.

Another important aspect of this estimate is the definition of the bins, specifically regarding both their number and their edges. Naively, one might partition the discriminant into bins of equal width. However, this approach poses the challenge of each bin having a very different number of events, which could result in less populated bins leading to erroneous conclusions due to statistical fluctuations. Moreover, we observe that partitioning the discriminant into too many bins exacerbates this issue; thus, the number of bins cannot be arbitrarily large relative to the given number of simulated (or expected) events.\footnote{Of course in the limit of infinite data this would produce a continuous approximation of the true distribution, but such limit is not possible in practice with a finite dataset.} We address the former problem by constructing bins based on (weighted) quantiles of the medium sample. This ensures that each bin has an equal expected yield, making them relatively robust against statistical fluctuations. With this solution, the latter problem is approached by seeking a regime where the vacuum fraction upper bound remains relatively stable against changes in the number of bins.

\begin{figure*}[h]
    \centering
    \includegraphics[scale=0.4]{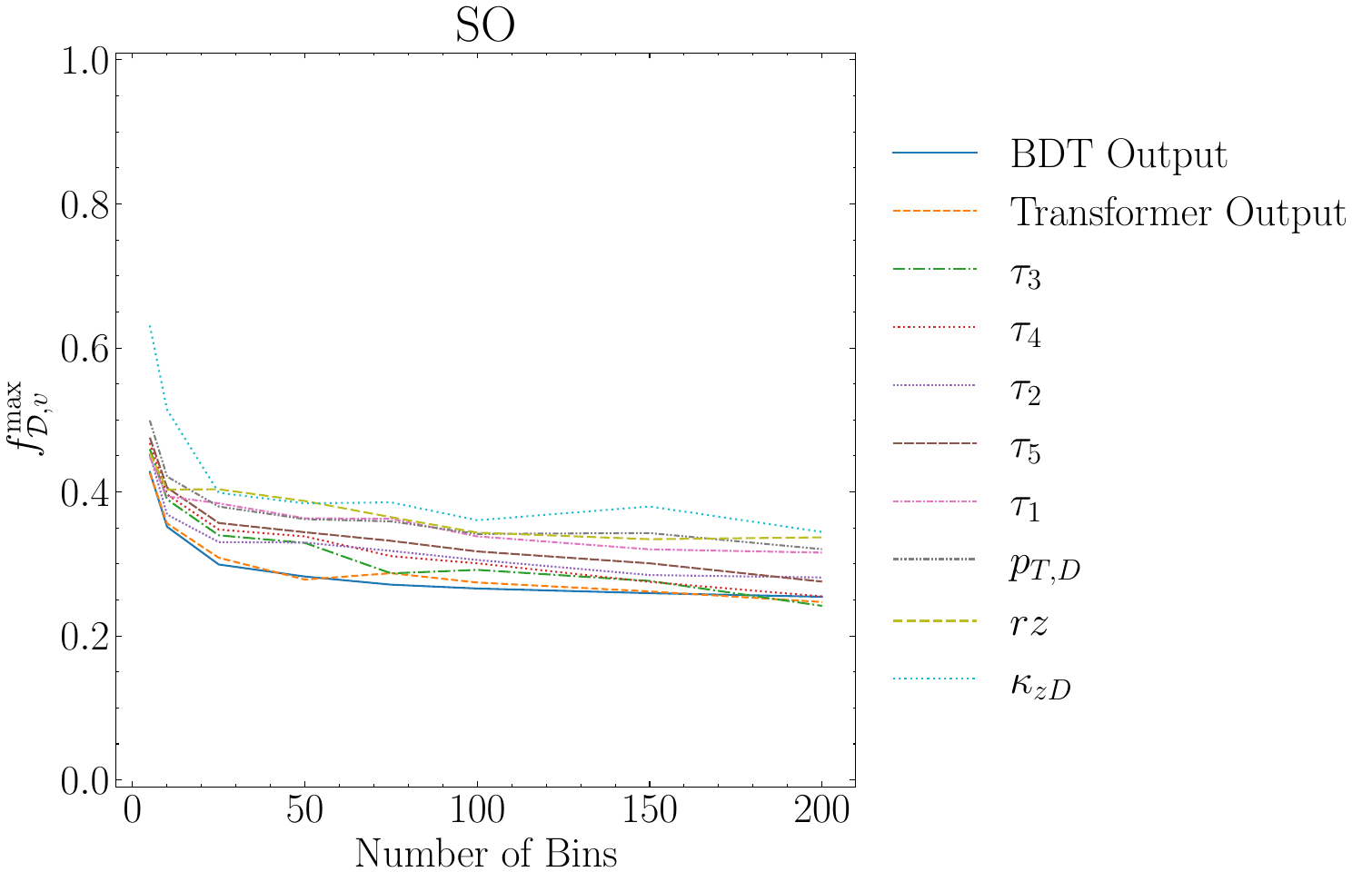}
    \includegraphics[scale=0.4]{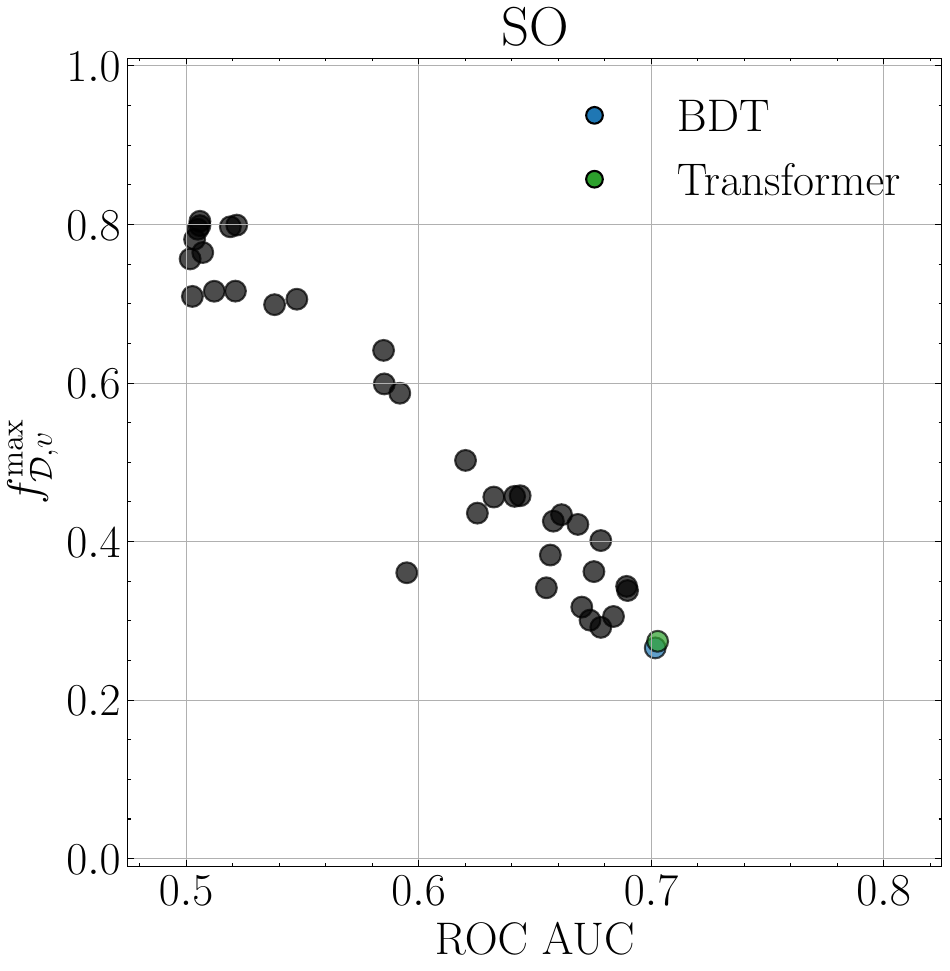}
    \caption{Left: the evolution of $f^{\text{max}}_{\mathcal{D},v}$ for the ten stringiest discriminants with the number of bins. Right: dependence of $f^{\text{max}}_{\mathcal{D},v}$ on the discriminant discriminating power, as measured by the ROC AUC, for 100 bins for all discriminants. Both plots for the SO scenario.}
    \label{fig:so-fractions}
\end{figure*}

In Fig.~\ref{fig:so-fractions} (left) we show how the upper bound on the fraction of vacuum changes with the number of bins for the ten discriminants. In addition to the Transformer and the BDT we consider the high-level observables from \cite{CrispimRomao:2023ssj} that produce the lowest upper bounds. We see that from $100$ bins onwards, the value of $f^{\text{max}}_{\mathcal{D},v}$ stabilises. On the right panel we fix the number of bins to be $100$ and show how $f^{\text{max}}_{\mathcal{D},v}$ gets smaller as the discriminating power of $\mathcal{D}$ increases, as expected.

\begin{figure*}[h]
    \centering
    \includegraphics[scale=0.4]{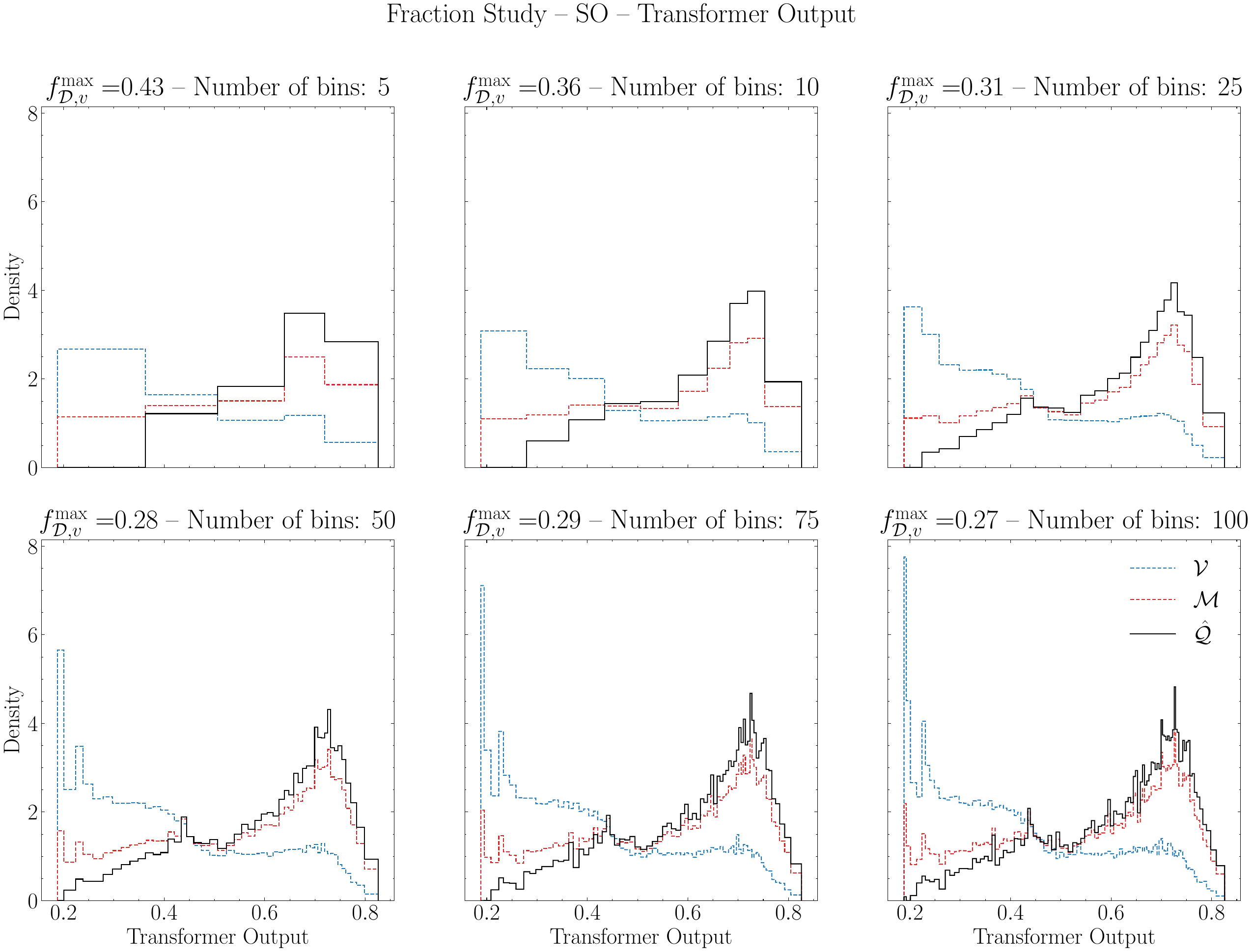}
    \caption{$f^{\text{max}}_{\mathcal{D},v}$ for the Transformer on the SO scenario for different number of bins. In red (blue) dashed lines we have the medium (vacuum) distributions. In black full line we have the estimated quenched distribution according to Eq.~\ref{eq:q-hat}.}
    \label{fig:so-fraction-transformer}
\end{figure*}

The results in Fig.~\ref{fig:so-fractions} indicate that the maximum fraction of vacuum-like jets in the medium sample is just under $0.3$. The discriminant that yields the strongest constraint is the BDT output, although it is very similar to that of the Transformer. In Fig.~\ref{fig:so-fraction-transformer} we visually illustrate this by defining the approximate distribution of quenched jets as
\begin{equation}\label{eq:q-hat}
    \hat Q_{\mathcal{D}}  =\frac{\mathcal{M}_{\mathcal{D}} - f^{\text{max}}_{\mathcal{D},v} \mathcal{V}_{\mathcal{D}}}{1-f^{\text{max}}_{\mathcal{D},v}}\ .
\end{equation}

\begin{figure*}[h]
    \centering
    \includegraphics[scale=0.4]{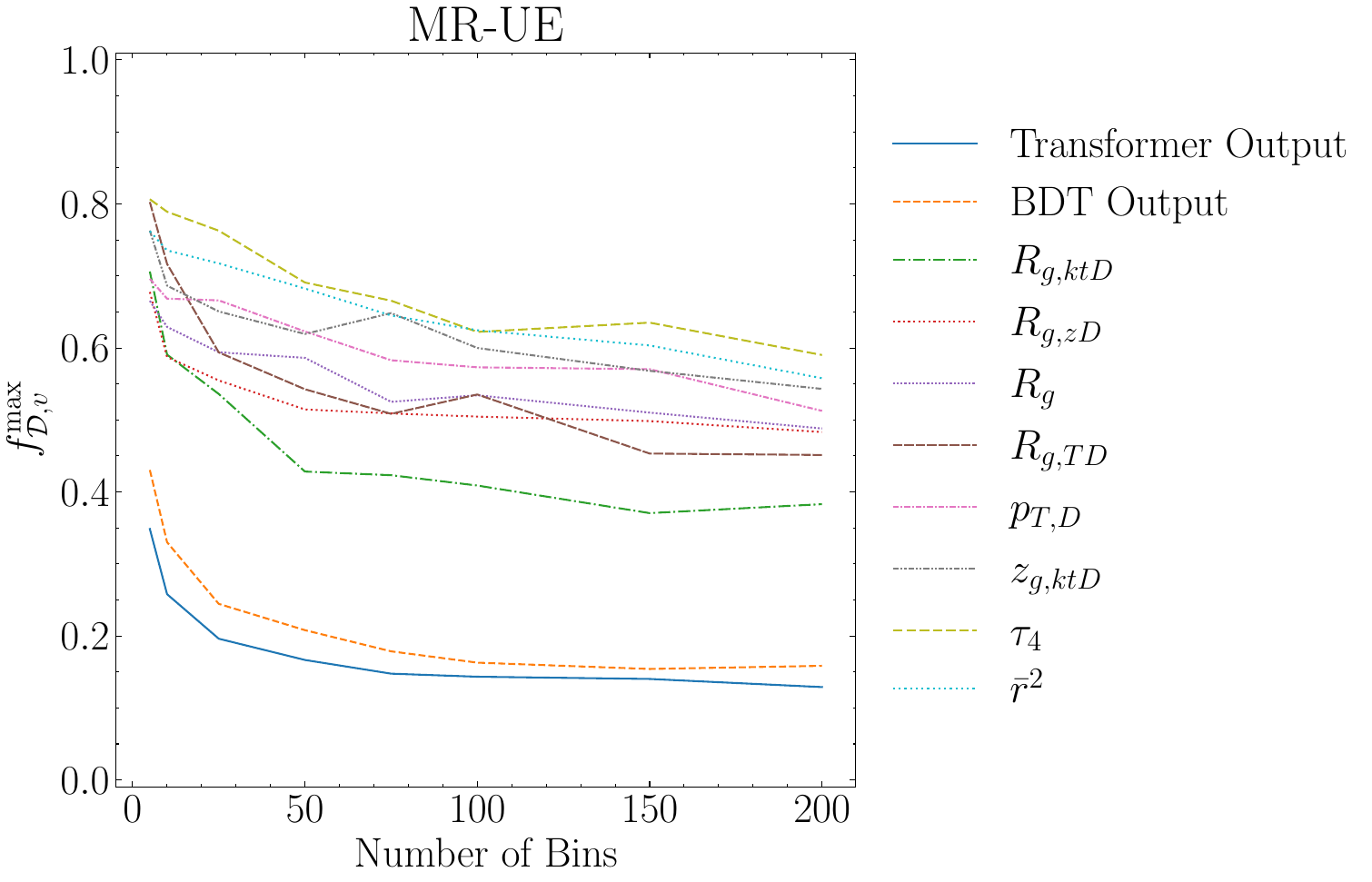}
    \includegraphics[scale=0.4]{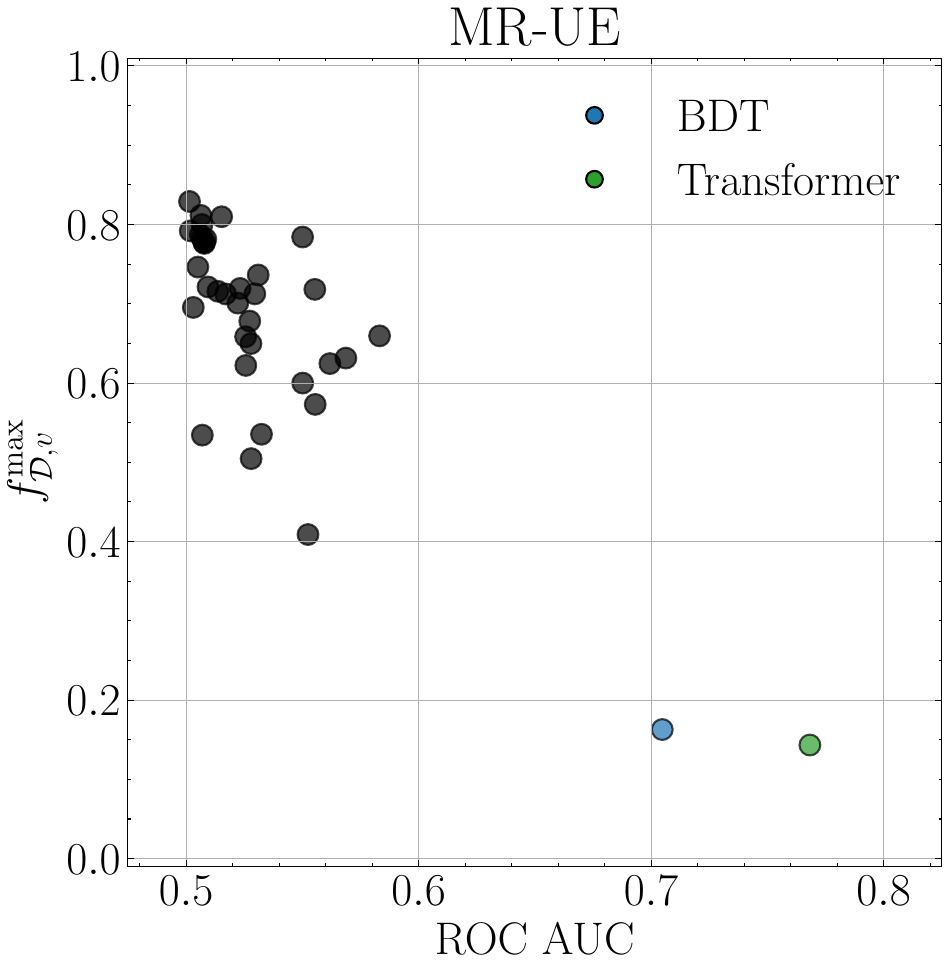}
     \caption{Left: the evolution of $f^{\text{max}}_{\mathcal{D},v}$ for the ten stringiest discriminants with the number of bins. Right: dependence of $f^{\text{max}}_{\mathcal{D},v}$ on the discriminant discriminating power, as measured by the ROC AUC, for 100 bins for all discriminants. Both plots for the MR-UE scenario.}
    \label{fig:mr-ue-fractions}
\end{figure*}

\begin{figure*}[h]
    \centering
    \includegraphics[scale=0.35]{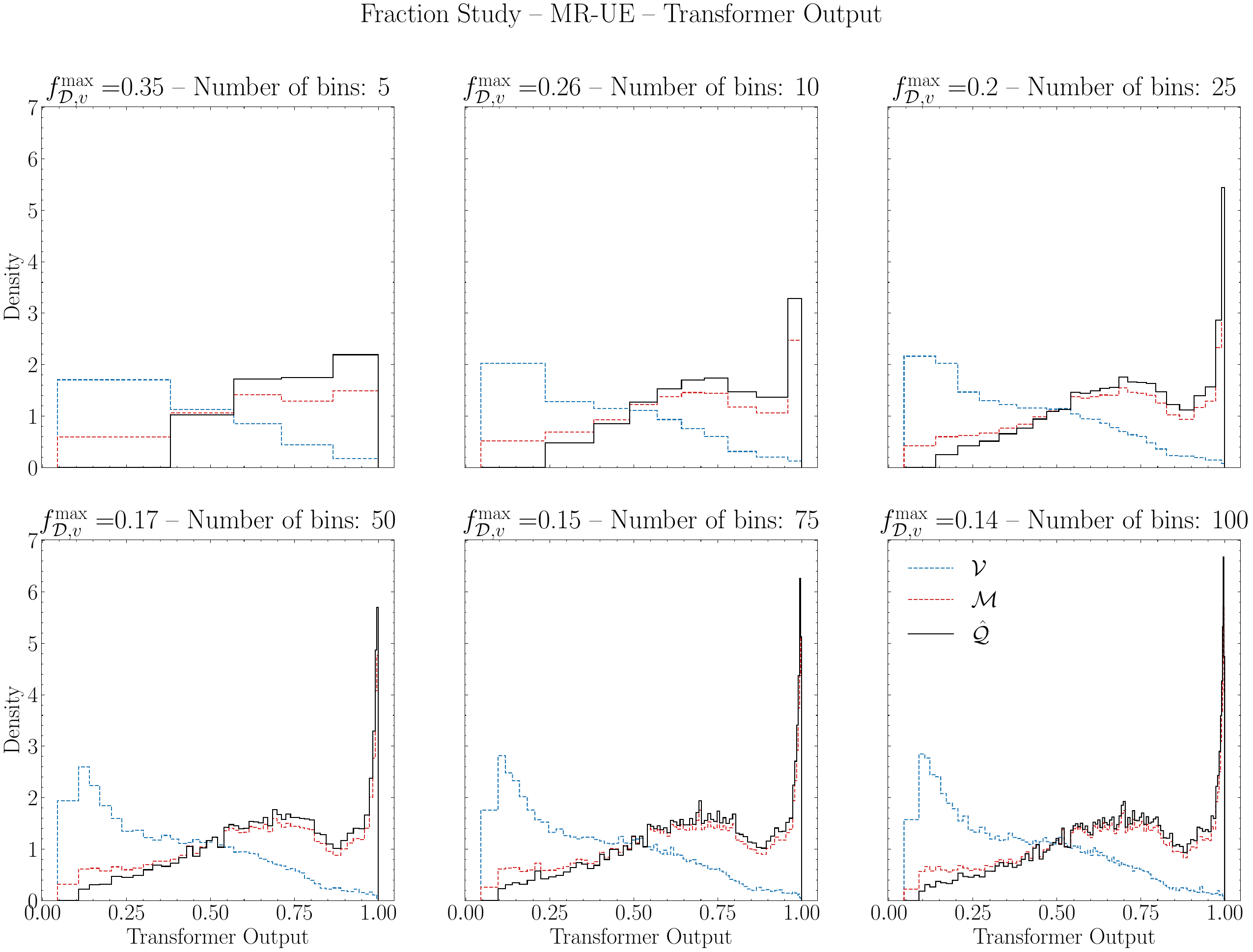}
    \caption{$f^{\text{max}}_{\mathcal{D},v}$ for the Transformer on the SO scenario for different number of bins. In red (blue) dashed lines we have the medium (vacuum) distributions. In black full line we have the estimated quenched distribution according to Eq.~\ref{eq:q-hat}.}
    \label{fig:mr-ue-fraction-transformer}
\end{figure*}

For the MR-UE case, medium response is a distinctive feature for jets in the medium sample. It is noteworthy that this feature is not captured by the high-level jet observables considered here. 
In Fig.~\ref{fig:mr-ue-fractions}, the left panel shows that both machine learning discriminants produce significantly stricter bounds on $f^{\text{max}}_{\mathcal{D},v}$ that, again, stabilise at around $100$ bins. We fix this number of bins for the right panel, where we show the relation between $f^{\text{max}}_{\mathcal{D},v}$ and the discrimination power of different observables. We observe a significant degradation, with respect to the SO case, of the single observable discriminating power with no single observable leading to a ROC AUC above 0.6. This degradation results on significantly poorer bounds for $f^{\text{max}}_{\mathcal{D},v}$ which now are now consistently less stringent that those obtained from both the BDT and the Transformer.
While the enhanced sensitivity of the Transformer to medium response leads to the most stringent bound on $f^{\text{max}}_{\mathcal{D},v}$ of just under $0.2$, this bound is not significantly better than yielded by the BDT. Both are significantly lower than that obtained in the SO case, confirming the role of medium response as a discriminating feature.
In Fig~\ref{fig:mr-ue-fraction-transformer} we can observe how the approximate quenching distribution emerges for different number of bins. Of especial relevance, we see the large accumulation of modified jets at the very end of the output, which are the jets that were not so confidently classified by the BDT. We see that this last bin is almost pure in jets from the medium sample, as previously seen in Fig~\ref{fig:mr-ue-transformer-vs-bdt}. We also note that, if we ignore the right-most bins, the distributions for vacuum and medium are very similar to the ones obtained for SO in Fig.~\ref{fig:so-fraction-transformer}.

\section{Conclusions}
\label{sec:conclusions}

The similar performance of our Transformer and the BDT, when QGP response and underlying event contamination are neglected, points towards the inability of the transformer to learn any  discriminating information beyond that captured by the high-level observables. While it is conceivable that such discriminating information is simply absent, we caution that several factors can limit the performance of the transformer and merit detailed study in future work: (i) transformer scaling laws \cite{kaplan2020scaling,henighan2020scaling,zhai2022scaling,hoffmann2022training} imply that very significant improvement could follow from a larger training data set; and (ii) our choice of parametrization for the 4-momentum of jet constituents, built very similarly to what is used to compute high-level observables, could introduce a bias and limit what the transformer can effectively learn. This limitation could be superseded by considering different representation bases or physics-aware architectures \cite{Spinner:2024hjm,Brehmer:2024yqw,Spinner:2025prg}.

When QGP response and underlying event are included, bringing our simulation results closer to experimental reality, the situation is qualitatively different. In this case, the presence of QGP response provides critical discriminating information, distinguishable from underlying event contamination, that is captured by the transformer and only margi\-nal\-ly by the BDT. 
We argued that this follows from the ability of the transformer to learn particle correlations beyond the 2-particle correlations that dominate the observables used to train the BDT. The discriminating power of multi-particle observables, in the form of Energy Flow Polynomials \cite{Komiske:2017aww}, has been explicitly verified recently \cite{Goncalves:2025asw}.
A salient feature of the classification performed by the transformer is the separation of the sample jets in PbPb into two clear classes: one that is unequivocally distinguishable from vacuum jets  (with outputs in the near vicinity of 1), and another with outputs with significant overlap with those  from pp (vacuum jets). This is directly attributable to the discriminating power of QGP response as the distribution of the transformer output for jets, not clearly identified as modified, is similar for both cases where QGP response is present and absent. 

The similarity between the fraction of vacuum-like jets in PbPb extracted from the Transformer and BDT outputs is worth comment. 
While this was to be expected when neglecting medium response and underlying event contamination as the outputs of the networks are closely correlated, the superior discriminating power of the transformer could be seen as implying a lower bound for the fraction of vacuum-like jets in the medium sample. The procedure we used to determine  $f^{\text{max}}_{\mathcal{D},v}$ in Eq.~\ref{eq:f_max} relies on the maximal excess of the vacuum output distribution with respect to the medium one. In other words, it relies on what the discriminants identify as vacuum-like jets. As the outputs of both the transformer and the BDT in the vicinity of $0$ are highly correlated even in the presence of QGP response and underlying event, see Fig.~\ref{fig:mr-ue-transformer-vs-bdt}, it is a natural that the inferred fraction of vacuum-like jets is similar in both cases.

\begin{acknowledgements}

This work is part of a project that has received funding from the European Research Council(ERC) under the European Union’s Horizon 2020 research and innovation programme (Grant agreement No. 835105, YoctoLHC). We acknowledge further support from Fundação para a Ciência e a Tecnologia (FCT), under ERC-PT A-Projects ‘Unveiling’, financed by PRR, NextGenerationEU (JGM) and under contract PRT/BD/151554/2021 (JAG). MCR is supported by the STFC under Grant No. ST/T001011/1.
\end{acknowledgements}

\appendix

\section{Jet Observables}
\label{app:observables}

For convenience, we briefly summarise the jet observables that are discussed in this work. A thorough study of these, and other, observables in the context of Vacuum-Medium discrimination was carried out in~\cite{CrispimRomao:2023ssj}, which we refer the reader for further details.

\paragraph{Angularities} The first set of observables is composed by generalise angularities~\cite{Larkoski:2014pca}, which are moments of the jet constituent distributions around the jet axis
\begin{equation}
    \lambda^\kappa_\beta = \sum_{i\in jet}z_i^\kappa R_{i,jet}^\beta \ ,
\end{equation}
where $z_i=p_{T,i}/p_{T,jet}$ is the fraction of the transverse momentum carried by the constituent $i$, and $R_{i,jet}$ is the angular distance to the jet axis, i.e.
\begin{equation}
    R_{i,j} =  \sqrt{(y_i - y_j)^2 + (\phi_i - \phi_j)^2} ,
\end{equation}
therefore, for $\beta\neq0$, the angularities measure the transverse distribution of jet constituents.

While $\kappa$ and $\beta$ can be any non-negative integer, the relevant angularities for this work are the momentum dispersion, $p_{T,D}$,~\cite{ALICE:2018dxf}, which reads
\begin{equation}
    p_{T,D} = \frac{\sqrt{\sum_{i\in jet} p^2_{T,i}}}{p_{T, jet}} = \sqrt{\lambda^2_0}\ ,
\end{equation}
and its mean over the number of jet constituents, \nconst,
\begin{equation}
    \bar z^2 = \frac{1}{\nconst} \lambda^2_0 = \frac{1}{\nconst} \sum_{i\in jet} \, z_{i}^2 \, .
\end{equation}

We notice that for $\kappa\neq1$ the generalised angularities are not IRC. Two IRC angularities appearing in this work are
\begin{align}
    rz &= \lambda^1_{1} \\
    r^2z &= \lambda^1_{2} \ . 
\end{align}

\paragraph{$N$-Subjettiness} Another set of observables relevant to this work are the $N$-Subjettiness observables~\cite{Thaler:2010tr}. These also capture the transverse properties of the jet and quantify how dissimilar is the jet from a collection of $N$ subjects. They read
\begin{equation}
    \tau_N = \frac{\sum_{i\in jet} p_T^i \min(R_{1,i}, \dots , \ R_{N,i} ) }{R_0 \;p_{T,jet}} \ ,
\end{equation}
with $R_0$ the jet clustering radius.

\paragraph{Grooming Derived} The final type of observables relevant to this work are those derived from grooming. For example, in SoftDrop,~\cite{Larkoski:2014wba}, the C/A reclustering jet branching history is recursively declustered, rejecting the softest branch, until the the following condition is met
\begin{equation}
    \frac{\min[p_{T,i}, p_{T,j}]}{p_{T,i} + p_{T,j}} > z_{cut} \left(\frac{R_{i,j}}{R_0}\right)^\beta \ ,
\end{equation}
where $i$,$j$ are the indices of the daughters of the branch, $R_{i,j}$ the angular distance between them, $R_0$ the jet radius of the initial clustering, and $z_{cut}$ and $\beta$ parameters governing the grooming. From this, one can derive three observables at the branching where the above condition is met. Let $p_{T,2}$ ($p_{T,1}$) be the softer (harder) branch at the branch splitting where the above condition is met, we then have
\begin{itemize}
    \item $z_g=p_{T,2}/(p_{T,1} + p_{T,2})$, i.e. the fraction of transverse momentum contained in the softer branch,
    \item $R_g=R_{1,2}$ the angular separation of two branches,
    \item $n_{SD}$ how many times the condition failed before it was met.
\end{itemize}

An alternative is to perform dynamical grooming,~\cite{Mehtar-Tani:2019rrk}, which selects the first C/A reclustering sequence branch satisfying
\begin{equation}
    \kappa^{(a)} = \frac{1}{p_{T,jet}} \max_{i \in \text{C/A seq}  } \left[ z_i ( 1 - z_i) p_{T,i} \left( \frac{R_{i,j}}{R_0} \right)^a \right],
\end{equation}
with $a$ a free parameter. Different values of $a$ capture different histories, 
\begin{itemize}
    \item TimeDrop (TD): $a=2$ 
    \item $k_T$-Drop (ktD): $a=1$ 
    \item $z$-Drop (zD): $a=0$ 
\end{itemize}
and we refer to~\cite{CrispimRomao:2023ssj,Mehtar-Tani:2019rrk} for more details. The relevant part for this work is that, for a given choice of $a$, one can derive the corresponding $R_g$, $z_g$, and $\kappa^{(a)}$ at the first branching that satisfies the grooming condition.

\section{Hyperparameter Optimisation}
\label{app:hp-opt}

The values for the hyperparameters were chosen using Bayesian optimisation implemented by Optuna~\cite{akiba2019optuna}, with the prior $N_h =1, \dots , 8$, $d_R = 4, \dots, 128$, $N_T = 1, \dots , 8$, and $M_h=0,\dots,2$, dropout rate in $[0.0.5]$, and maximal learning rate in $\{10^{-5},3\times10^{-5},10^{-4},3\times10^{-4},10^{-3},3\times10^{-3}\}$. The remaining hyperparamters were set to default. We used \texttt{Adam} optimizer with a cosine annealing learning rate with warm-up set to the first 10 epochs. The classifier was trained against the binary cross-entropy loss function between the vacuum and the medium classes, with early stop if no improvement was seen after 50 epochs. When stopped, the weights of the best epoch were restored and the Area Under the Curve of the Receiver Operating Characteristic (ROC AUC) was computed on the validation set. After 100 hyperparameter optimisation trials, we trained separately for the SO and the MR-UE physics cases. In both cases, the Monte Carlo generation statistical weights returned by \jewel\ were used during training to ensure the correct statistics of the loss function, with an overall class reweighing imposed to guarantee the same statistical contribution from both classes during training.

The best combination of hyperparameters for the Transformer trained on the SO dataset was $N_h=3$, $d_R=5$, $N_T=3$, $M_h=1$, dropout rate at $0.3$, maximal learning rate of $10^{-3}$, and attention pooling. For the MR-UE case, the best hyperparameters were $N_h=1$, $d_R=109$, $N_T=3$, $M_h=1$, dropout rate at $0.2$, maximal learning rate of $3\times10^{-4}$, and attention pooling. 
The hyperparameters of the BDT were also optimised to guarantee maximal performance, but the details of this step are omitted.


\bibliographystyle{spphys}       

\bibliography{paper}

\end{document}